\documentclass[12pt]{article}
\usepackage{amssymb}
\usepackage{amsfonts}
\usepackage{amsmath}

\newtheorem{theorem}{Theorem}

\newtheorem{proposition}[theorem]{Proposition}

\begin{document}

\title{Polymer Measure: Varadhan's Renormalization Revisited}
\author{\textbf{Wolfgang Bock} \\
{\small CMAF, University of Lisbon, P 1649-003 Lisbon, Portugal}\\
{\small bock@campus.ul.pt} \and \textbf{Maria Jo\~{a}o Oliveira} \\
{\small Universidade Aberta, P 1269-001 Lisbon, Portugal}\\
{\small CMAF, University of Lisbon, P 1649-003 Lisbon, Portugal}\\
{\small oliveira@cii.fc.ul.pt} \and \textbf{Jos\'e Lu{\'\i}s da Silva} \\
{\small CCEE, University of Madeira, P 9000-390 Funchal, Portugal}\\
{\small CCM, University of Madeira, P 9000-390 Funchal, Portugal}\\
{\small luis@uma.pt} \and \textbf{Ludwig Streit} \\
{\small Forschungszentrum BiBoS, Bielefeld University, D 33501 Bielefeld,
Germany}\\
{\small CCM, University of Madeira, P 9000-390 Funchal, Portugal}\\
{\small streit@physik.uni-bielefeld.de}}
\date{}
\maketitle

\begin{abstract}
Through chaos decomposition we improve the Varadhan estimate for the rate of
convergence of the centered approximate self-intersection local time of
planar Brownian motion.
\end{abstract}

\noindent \textbf{Keywords:} Edwards model, self-intersection local time,
Varadhan renormalization, white noise analysis

\smallskip

\noindent \textbf{Mathematics Subject Classifications (2010):} 28C20, 41A25,
60H40, 60J55, 60J65, 82D60

\section{Introduction}

The Edwards model \cite{Edwards} for self-repelling or "weakly
self-avoiding" $d$-dimensional Brownian motion, with applications in polymer
physics and quantum field theory, is informally given by a Gibbs factor 
\begin{equation*}
G=\frac{1}{Z}\exp \left( -g\int_{0}^{T}ds\int_{0}^{t}dt\delta \left(
B(s)-B(t)\text{ }\right) \right) 
\end{equation*}%
with $g>0$ and$\ \ $ 
\begin{equation*}
Z=E\left( \exp \left( -g\int_{0}^{T}ds\int_{0}^{t}dt\delta \left( B(s)-B(t)%
\text{ }\right) \right) \right) .
\end{equation*}%
Using%
\begin{equation*}
\delta _{\varepsilon }(x):=\frac{1}{(2\pi \varepsilon )^{d/2}}e^{-\frac{%
|x|^{2}}{2\varepsilon }},\quad \varepsilon >0,
\end{equation*}%
one defines an approximate self-intersection local time by 
\begin{equation*}
L_{\varepsilon }:=\int_{0}^{T}dt\int_{0}^{t}ds\,\delta _{\varepsilon
}(B(t)-B(s)).
\end{equation*}%
For $d\geq 2$%
\begin{equation*}
\lim_{\varepsilon \searrow 0}\mathbb{E}(L_{\varepsilon })=\infty .
\end{equation*}%
For the planar case $d=2,$ Varadhan \cite{v} has shown that centering 
\begin{equation*}
L_{\varepsilon ,c}\ :=L_{\varepsilon }\ -\mathbb{E}(L_{\varepsilon })\ \text{%
and}\ L_{c}:=\lim_{\varepsilon \searrow 0}L_{\varepsilon ,c}
\end{equation*}%
is sufficient to make the Gibbs factor $G=Z^{-1}\exp \left( -gL_{c}\right) $
well defined. An estimate for the rate of convergence%
\begin{equation*}
\Vert L_{c}\ -L_{\varepsilon ,c}\Vert _{2\ }^{2}\leq const.\varepsilon
^{\alpha }
\end{equation*}%
\ for all $\alpha <1/2$, is in Varadhan's words, "the most difficult step of
all and requires considerable estimation". In this note we shall use a
multiple Wiener integral or chaos expansion for an alternate and
comparatively straightforward argument, extending the estimate to all $%
\alpha <1$.

\section{Fock space representation of the local time}

\label{Section1}

The Ito-Segal-Wiener isomorphism relates the $L^{2}\,$\ space of planar
Brownian motion with the Fock space%
\begin{equation*}
\mathfrak{F}=\left( \bigoplus_{n=0}^{\infty }\mathrm{Sym}\,L^{2}(\mathbb{R}%
^{n},n!d^{n}x)\right) ^{\otimes 2}.
\end{equation*}%
We shall use the multi-index notation 
\begin{equation*}
\mathbf{n}=(n_{1},n_{2})\ \quad n=n_{1}+n_{2},\quad \mathbf{n}!=n_{1}!n_{2}!
\end{equation*}%
The Fock space norm is then%
\begin{equation*}
\Vert F\Vert _{2}^{2}=\sum_{n_{i}\geq 0}\mathbf{n}!\left\Vert F_{\mathbf{n}%
}\right\Vert _{2}^{2}.
\end{equation*}
For $L_{\varepsilon ,c}$ the kernel functions $F_{\mathbf{n}}$ were
computed explicitly in \cite{fhsw}. For the planar case the result is

\begin{proposition}
\cite{fhsw}: For $d=2$ the kernel functions $F_{\mathbf{n}}$ of 
$L_{\varepsilon ,c}(T)$ and $L_{c}(T)$ have their support on $\left[ 0,T\right] ^{n}$ 
and are, with $\varepsilon >0$, and $\varepsilon =0$ respectively, for $n>1$ 
\begin{align*}
& F_{2\mathbf{n},\varepsilon }(u_{1},\ldots ,u_{2n})=\frac{1}{2\pi }\left( -%
\frac{1}{2}\right) ^{n}\frac{1}{n(n-1)\,\mathbf{n!}} \\
& \times \left( \frac{1}{(T+\varepsilon )^{n-1}}-\frac{1}{(v+\varepsilon
)^{n-1}}-\frac{1}{(T-u+\varepsilon )^{n-1}}+\frac{1}{(v-u+\varepsilon )^{n-1}%
}\right) ,
\end{align*}%
where $v:=\max_{1\leq k\leq 2n}u_{k}\leq T$ and $u:=\min_{1\leq k\leq 2n}u_{k}\geq 0$. For $n=1$
\begin{equation*}
F_{2,\varepsilon }(u_{1},u_{2})=-\frac{1}{4\pi }\left( \ln (v+\varepsilon
)+\ln (T-u+\varepsilon )-\ln (v-u+\varepsilon )-\ln (T+\varepsilon )\right) .
\end{equation*}%
All kernel functions $F_{\mathbf{n}}$ with odd $n_{i}$ are zero.
\end{proposition}

\subsection{The rate of convergence}

\label{Section2}

\begin{theorem}
\label{ratetrunc} Given $T>0$. Then for any $\alpha <1$ there is a constant $%
C_{T,\alpha }>0$ such that for all $\varepsilon >0$ 
\begin{equation*}
\Vert L_{\varepsilon ,c}(T)-L_{c}(T)\Vert _{2}^{2}\leq C_{T,\alpha
}\varepsilon ^{\alpha }.
\end{equation*}
\end{theorem}

\noindent \textbf{Proof:} From Proposition 1 
\begin{equation*}
\left\Vert F_{2\mathbf{n},0}-F_{2\mathbf{n},\varepsilon }\right\Vert
_{2}^{2}=\Big(n(n-1)2\pi 2^{n}\mathbf{n}!\Big)^{-2}%
\int_{0}^{T}d^{2n}u_{k}K_{\varepsilon }^{2}(u,v,T)
\end{equation*}%
where for $n>1$%
\begin{eqnarray*}
K_{\varepsilon }(u,v,T) &=&\Big(T^{-n+1}-(T+\varepsilon )^{-n+1}\Big)-\Big(%
v^{-n+1}-(v+\varepsilon )^{-n+1}\Big) \\
&&-\Big((T-u)^{-n+1}-(T-u+\varepsilon )^{-n+1}\Big)+\Big((v-u)^{-n+1}-(v-u+%
\varepsilon )^{-n+1}\Big).
\end{eqnarray*}%
Since $K_{\varepsilon }(u,v,T)$ does not depend on $2n-2$ of the $u_{k}$%
-variables, we may integrate them out: 
\begin{equation*}
\left\Vert F_{2\mathbf{n},0}-F_{2\mathbf{n},\varepsilon }\right\Vert
_{2}^{2}=\Big(n(n-1)2\pi 2^{n}\mathbf{n}!\Big)^{-2}2n(2n-1)\int_{0}^{T}dv%
\int_{0}^{v}du(v-u)^{2n-2}K_{\varepsilon }^{2}(u,v,T)\,.
\end{equation*}%
Of the four terms in $K_{\varepsilon }$, the last one is dominant so that

\begin{eqnarray}
\left\Vert F_{2\mathbf{n},0}-F_{2\mathbf{n},\varepsilon }\right\Vert
_{2}^{2} &\leq &16\frac{2n(2n-1)}{\Big(n(n-1)2\pi 2^{n}\mathbf{n}!\Big)^{2}}%
\int_{0}^{T}dv\int_{0}^{v}du\left( 1-\left( \frac{v-u}{v-u+\varepsilon }%
\right) ^{n-1}\right) ^{2}  \notag \\
&=&16\frac{2n(2n-1)}{\Big(n(n-1)2\pi 2^{n}\mathbf{n}!\Big)^{2}}%
\int_{0}^{T}d\tau \int_{0}^{T-\tau }du\left( 1-\left( \frac{\tau }{\tau
+\varepsilon }\right) ^{n-1}\right) ^{2}  \notag \\
&\leq &16\frac{2n(2n-1)T}{\Big(n(n-1)2\pi 2^{n}\mathbf{n}!\Big)^{2}}%
\int_{0}^{T}d\tau \left( 1-\left( \frac{\tau }{\tau +\varepsilon }\right)
^{n-1}\right) ^{2}  \notag \\
&=&16\frac{2n(2n-1)T}{\Big(n(n-1)2\pi 2^{n}\mathbf{n}!\Big)^{2}}%
\int_{0}^{T}d\tau \left( \tau ^{n-1}(n-1)\int_{0}^{\varepsilon }\,\frac{dx}{%
(x+\tau )^{n}}\right) ^{2}\,  \label{est}
\end{eqnarray}

By H\"older's inequality%
\begin{equation*}
\int_{0}^{\varepsilon }\frac{dx}{(x+\tau )^{n}}\leq \varepsilon ^{1/q}\left(
\int_{0}^{\infty }\frac{dx}{(x+\tau )^{np}}\,\right) ^{1/p}=\varepsilon
^{1/q}\left( \frac{1}{np-1}\tau ^{1-np}\right) ^{1/p}
\end{equation*}%
if $\frac{1}{q}+\frac{1}{p}=1$. Insertion of this estimate into (\ref{est})
produces 
\begin{equation*}
\sum_{\mathbf{n:}n>1}(2\mathbf{n})!\left\Vert F_{2\mathbf{n},0}-F_{2\mathbf{n%
},\varepsilon }\right\Vert _{2}^{2}\leq \frac{4p}{(2-p)\pi ^{2}}T^{\frac{2}{p%
}}\varepsilon ^{\frac{2}{q}}\sum_{\mathbf{n:}n>1}(2\mathbf{n})!\frac{2n(2n-1)%
}{(n2^{n}\mathbf{n}!)^{2}}\frac{1}{(pn-1)^{\frac{2}{p}}}
\end{equation*}%
which is convergent if $\frac{2}{p}>1$, i.e., $q>2$. For the $n=1$ term an $%
\varepsilon ^{\frac{2}{q}}$ \ estimate is likewise obtained via H\"older's
inequality.

Hence, for any $\alpha <1$,%
\begin{equation*}
\Vert L_{c}(T)-L_{\varepsilon ,c}(T)\Vert _{2}^{2}=\sum_{\mathbf{n:}n\geq 1}(2%
\mathbf{n})!\left\Vert F_{2\mathbf{n},0}-F_{2\mathbf{n},\varepsilon
}\right\Vert _{2}^{2}\leq C(T,\alpha )\varepsilon ^{\alpha },\quad \forall
\varepsilon >0.
\end{equation*}%
$\hfill\blacksquare $

\subsection*{Acknowledgments}

Financial support of FCT through the research project PTDC/MAT-STA/1284/2012
is gratefully acknowledged. W.~B.~and J.~L.~S.~also would like to thank for
financial support of IGK and DFG through SFB-701 (University of Bielefeld),
respectively.

\end{document}